\documentclass[pre,twocolumn]{revtex4}
\usepackage{amsfonts}
\usepackage{amsmath}
\usepackage{amssymb}
\usepackage{graphicx}

\begin{document}

\title{Role of boundary constraints in DNA cyclization}
\author{Alexei V. Tkachenko}
\affiliation{Department of Physics and Michigan Center for Theoretical Physics,
University of Michigan, 450 Church Str., Ann Arbor, 48109 MI, USA}

\begin{abstract}
We modify the classical Shimada-Yamakawa theory of DNA looping by
generalizing the form of boundary constraints. This generalization is
important in the context of DNA cyclization experiments since it mimics the
reduced local rigidity of the "nicked" DNA loop. Our results indicate that
the non-trivial boundary constraints may be responsible for the existing
dramatic discrepancy between various DNA cyclization experiments. The
developed effective Hamiltonian method may be extended to even broader class
of DNA looping problems.

\end{abstract}

\maketitle

Loop formation in double-stranded DNA (dsDNA) is essential for such
important biological processes as regulation of gene expression and DNA
packaging into nucleosomes. \ In the first case, the loop is induced by the
interactions between transcription factor proteins bound to different sites
along the DNA chain \cite{loop-review}-\cite{lacflex}. In the case of
nucleosome, DNA wraps around the near-cylindrical histone octamer. \ The
classical theory of looping, based on the elastic description of DNA, was
proposed more than two decades ago by Shimada and Yamakawa (SY) \cite%
{wlc-hairpin1}. There are however multiple indications that the original
theory is not sufficiently adequate for describing the real experimental
situation.

\begin{figure}[tbp]
\begin{center}
\includegraphics[
height=1.8in,
width=3.5in
]
{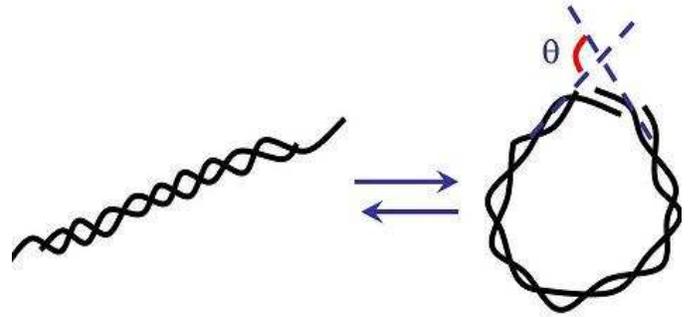}
\end{center}
\caption{Scheme of DNA cyclization experiment. Looping is caused by
hibridization of mutually complementary ssDNA ends. }
\label{loop_pict}
\end{figure}

Partially, the deviations can be attributed to the complexity of the actual 
\textit{\ in vivo} problem. However, a large discrepancy is also reported in
recent \textit{in vitro} experiments on DNA cyclization \cite{Widom1}\cite%
{Widom2}. In these experiment, looping is induced by hybridization of
mutually complementary ssDNA ends of the chain (see Figure \ref{loop_pict}).
Even in this relatively simple case, the agreement between theory and
experiment is a highly controversial issue.

On the one hand, older cyclization experiments by Shore et al do agree with
SY model \cite{ring1}\cite{ring2}. In fact, those data were used in the
original SY paper to support their model. On the other hand, Cloutier and
Widom reported that the discrepancy between experimental looping probability
and SY results may reach two to three orders of magnitude \cite{Widom1}.
Their paper was followed by the work of Vologodskii lab, in which the
agreement between theory and experiment was confirmed \cite{volo}. More
recently, Cloutier and Widom reiterated their claim, and also reported that
the twist-related oscillations of the looping probability are strongly
reduced compared to the SY prediction\ \cite{Widom2}. This controversy
inspired a new interest to the problem among theorists. In particular, it
was suggested that strong bending may induce local structural defects such
as "kinks" \ or ssDNA "bubbles" (which act as "soft" kinks)\cite{bubble}-%
\cite{bubble-ind}. At present, there is no direct evidence to support any of
these models, and they do not resolve the conflict between the different
experiments.

In this communication, we propose an alternative explanation to the existing
controversy. We argue that the DNA cyclization involve more complicated
boundary (i.e. terminal) constraints then it is traditionally believed. In
order to solve the problem with the modified boundary conditions, we develop
an effective Hamiltonian method. The approach is potentially applicable
beyond the scope of this work, e.g. for study of protein-mediated looping 
\cite{cocco1}.

The SY theory of looping starts with dsDNA modelled as an elastic rod
subjecxted to thermal fluctuations, which is essensially equivalient to Worm
Like Chain (WLC) model \cite{KP}-\cite{Marko}:

\begin{equation}
H_{0}=k_{B}T\int\limits_{0}^{L}ds\left[ \frac{l_{p}}{2}\left( \frac{\partial 
\mathbf{\hat{t}}}{\partial s}\right) ^{2}+\frac{l_{t}}{2}\left( \frac{%
\partial \mathbf{\psi }}{\partial s}\right) ^{2}\right]
\end{equation}%
Here, the first term describes the bending elasticity with modulus
proportional to dsDNA persistence length $l_{p}\approx 50nm$,\ the second
term represents torsional elasticity, with the modulus proportional to twist
persistence length, $l_{t}\simeq 100nm$. $\mathbf{\hat{t}}$ is unit tangent
vector of the chain, and $\mathbf{\psi }$ is twist angle, both functions of
position along the chain, $s$.

From the experimental and biological points of view, the most important
characteristics of the looping problem is the so called $J$-factor. It has a
meaning of an effective concentration of one end of the loop in the vicinity
of the other, at the open configuration. $J$-factor can be related to
equilibrium probability for the loop to be closed, $P^{\left( loop\right) }$%
, as well as to the closing/opening times ($\tau _{cl}$, $\tau _{op}$):%
\begin{equation}
\frac{P^{\left( loop\right) }}{1-P^{\left( loop\right) }}=\frac{\tau _{op}}{%
\tau _{cl}}=\frac{J}{c_{0}}\exp \left( \frac{\varepsilon }{k_{B}T}\right)
\end{equation}%
Here $\varepsilon $ is the mutual affinity of the loop terminals, and $%
c_{0}=1M$ is the standard reference concentration. Affinity $\varepsilon $
can be independently determined from the experiment on dimerization of free
reacting terminal groups in solution.

Shimada and Yamakawa calculated $J$-factors for two important cases:
circular loops with completely aligned ends ($\mathbf{\hat{t}}\left(
L\right) =\mathbf{\hat{t}}\left( 0\right) $):%
\begin{equation}
J_{0}\approx \frac{32\pi ^{3}}{l_{P}^{3}}\left( \frac{l_{P}}{L}\right)
^{6}\exp \left( -\frac{2\pi ^{2}l_{p}}{L}-\frac{L}{4l_{P}}\right) ,
\label{J0}
\end{equation}%
and loops with unconstrained orientations of the end segments:%
\begin{equation}
J_{free}\approx \frac{110}{l_{P}^{3}}\left( \frac{l_{P}}{L}\right) ^{5}\exp
\left( -\frac{14.04l_{p}}{L}-\frac{L}{4l_{P}}\right) .  \label{J1}
\end{equation}%
These results are obtained in the limit of short loops. However, their range
of applicability extends up to $L\sim 10l_{P}$ (i.e. 1500 bp). The effects
of torsional constraints which are not included in Eqs.(\ref{J0})-(\ref{J1}%
), result in an additional factor with a pronounced oscillatory behavior
with the period of DNA helical turn, 10.5 bp.

DNA loop formed in a reversible cyclization experiments (i.e. before
ligation) is not identical to a circular DNA. It is "nicked" in two points
corresponding to the ends of the DNA strands (see Figure \ref{loop_pict}).
These singular points are expected to have a greater flexibility then the
rest of the chain \cite{F-K}. The effective local flexibility must strongly
depend on the base-stacking interactions. We will describe the coupling
between orientations of the end segments of the loop with the following
minimal model: 
\begin{equation}
\frac{H_{end}}{kT}=\frac{\kappa \theta ^{2}}{2}+\frac{\kappa ^{\prime
}\left( \Delta \psi \right) ^{2}}{2}
\end{equation}%
Here $\theta $ is the angle between directions of tangent vectors, $\mathbf{%
\hat{t}}\left( 0\right) $ and $\mathbf{\hat{t}}\left( L\right) $, and $%
\Delta \psi $ is the relative twist of the two segments. The two parameters, 
$\kappa $ and $\kappa ^{\prime }$ should be of the same order for the given
sequence. Since the characteristic stacking energy is of order of $kT$, we
expect $\kappa \simeq \kappa ^{\prime }\sim 1$.

In order to calculate $J-$ factor with this modified boundary constraints,
we develop an \textit{effective Hamiltonian method} (EHM) for the problem.
Effective Hamiltonian can be introduced as a free energy of the loop
parameterized with orientations of its end segments. If we neglect any
torsional constraints (i.e. assume $\kappa ^{\prime }=0$), the Hamiltonian
of a closed DNA loop can be written as a function of azimuthal and polar
angles, $\theta $ and $\varphi $ of tangent vector $\mathbf{\hat{t}}\left(
L\right) $, with respect to the vector triad at the other end of the loop, $%
s=0$:

\begin{equation}
H=H_{loop}\left( \theta ,\varphi \right) +\frac{\kappa \theta ^{2}}{2}
\label{Ham}
\end{equation}

\begin{figure}[ptbh]
\begin{center}
\includegraphics[
height=2.527in,
width=3.3607in
]{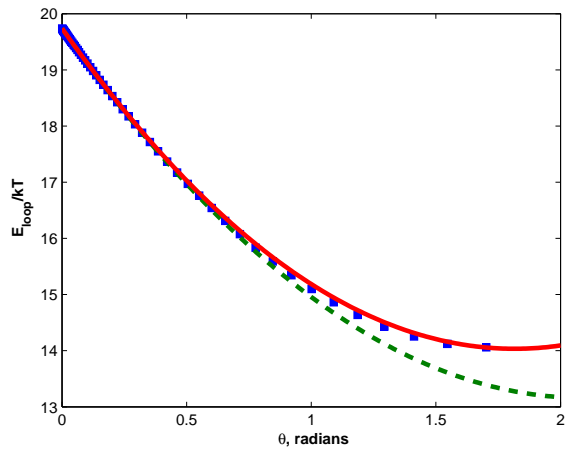}
\end{center}
\caption{Comparison of the Effective Hamiltonian \ (solid line), to the
exact \ result for the elastic energy of a loop (squares). Dashed line shows
the extrapolation from point $\protect\theta=0$, without any adgustment to
model parameter $\protect\gamma$ (i.e $\protect\gamma=3$). }
\label{loop_enrg}
\end{figure}

In order to construct $H_{loop}\left( \theta ,\varphi \right) $, we first
discuss the ground state energy of the loop. Since the original Hamiltonian
of WLC model is formally equivalent to Lagrangian of symmetrical spinning
top, finding its ground state is an integrable mechanical model. However,
finding the exact solution for a particular set of boundary conditions,
requires an inversion of incomplete Elliptical functions, and therefore it
is not practical \cite{cocco1}. Instead, we consider the problem in the
vicinity of circular loop configuration (which corresponds to $\mathbf{\hat{t%
}}\left( 0\right) =\mathbf{\hat{t}}\left( L\right) $). In this limit, the \
ground state energy of a 2D (planar) loop can be written as an expansion in $%
\theta $:

\begin{equation}
\frac{E_{loop}\left( \theta \right) }{kT}\approx \frac{l_{P}}{L}\left( 2\pi
^{2}+\beta \theta +\frac{\gamma \theta ^{2}}{2}\right) +O\left( \theta
^{3}\right) ,  \label{eloop}
\end{equation}%
where the exact values of the coefficients are: $\beta =2\pi $; $\gamma =3$.
Already this expression gives an excellent description of the global
behavior of function $E_{loop}\left( \theta \right) $, not limited to the
near vicinity of circular loop ($\theta =0$ ). In particular, it predicts
the minimal energy, $E_{\min }/kT=(4\pi ^{2}/3)l_{p}/L$ $\approx
13.16l_{p}/L $, which is very close to the results of complete numerical
solution of the problem: $E_{\min }/kT$ $\approx 14.04l_{p}/L$.
Nevertheless, due to the strong exponential dependence of $J$-factor on the
elastic energy, we need a further improvement of the analytic formula for $%
E_{loop}\left( \theta \right) $. We achieve this by treating the coefficient 
$\gamma $ in as a free parameter, and adjusting it to value $\gamma =3.46$.
This leads to an exact matching of the minimum energy. The overall behavior
of approximate expression, Eq.(\ref{eloop}), becomes nearly
indistinguishable from the exact result (see Figure \ref{loop_enrg}).

Since the change in the elastic energy is the dominant correction to the
loop free energy, we can write the effective Hamiltonian of the loop in the
following form:

\begin{equation}
\frac{H_{loop}\left( \theta,\varphi\right) }{kT}\approx\frac{F_{0}}{kT}+%
\frac{l_{P}}{L}\left( 2\pi\theta\cos\varphi+\frac{\left( \gamma
+\gamma^{\prime}\sin^{2}\varphi\right) \theta^{2}}{2}\right) ,\text{ \ }
\label{Heff}
\end{equation}
Here, the earlier expression, Eq. (\ref{eloop}), has been generalized for
the case of 3D loops, and additional fine-tuning parameter, $\gamma^{\prime}$
has been introduced. $F_{0}$ is free energy of the circular loop.

The combined effective Hamiltonian Eq. (\ref{Ham}), reaches minimum at \ $%
\theta=\theta^{\ast}=-2\pi/\left( \gamma+\kappa L/l_{p}\right) $, and $%
\varphi=0$. We now expand it in the vicinity of that point:

\begin{equation}
\frac{H\left( \theta ,\varphi \right) }{kT}\approx \frac{F_{0}}{kT}+\frac{%
2\pi ^{2}l_{P}}{L\left( \gamma +\kappa L/l_{p}\right) }+\left( \frac{\gamma
l_{P}}{L}+\kappa \right) \frac{\left( \theta -\theta ^{\ast }\right) ^{2}}{2}%
+
\end{equation}%
\begin{equation*}
+\frac{4\pi ^{2}l_{P}}{L\left( \gamma +\kappa L/l_{p}\right) }\left[ 1+\frac{%
\gamma ^{\prime }}{\left( \gamma +\kappa L/l_{p}\right) }\right] \frac{%
\varphi ^{2}}{2}
\end{equation*}%
This quadratic expansion allows one to calculate the overall free energy of
the looped state in Gaussian approximation, and hence obtain $J$-factor for
arbitrary coupling $\kappa $ :%
\begin{equation}
J_{\kappa }=J_{0}\frac{\int \mathrm{e}^{-\left( H-F_{0}\right) /kT}d\Omega }{%
\int \mathrm{e}^{-H_{end}/kT}d\Omega }
\end{equation}

The result of this calculation can be well approximated by the following
analytic expression (the torsional effects are omitted): 
\begin{equation}
J_{\kappa}\approx J_{0}\left( L/l_{p}\right) \frac{\left( 2\kappa+1\right) L%
}{4\pi l_{p}}\sqrt{1+\frac{1}{1+2\kappa L/\gamma l_{p}}}\times  \label{Jk}
\end{equation}%
\begin{equation*}
\times\sin\left( \frac{2\pi}{\gamma+\kappa L/l_{p}}\right) \exp\left( \frac{%
2\pi^{2}}{\left( \gamma+\kappa L/l_{p}\right) }\frac{l_{P}}{L}\right)
\end{equation*}

This form generalizes of the original \ SY result \ The parameter $%
\gamma^{\prime}$ was tuned to the value $\gamma^{\prime}=-\gamma/2$, to
achieve a nearly exact matching with both SY limits, Eqs. (\ref{J0}) and (%
\ref{J1}).

\begin{figure}[ptbh]
\begin{center}
\includegraphics[
height=2.6437in,
width=3.5163in
]{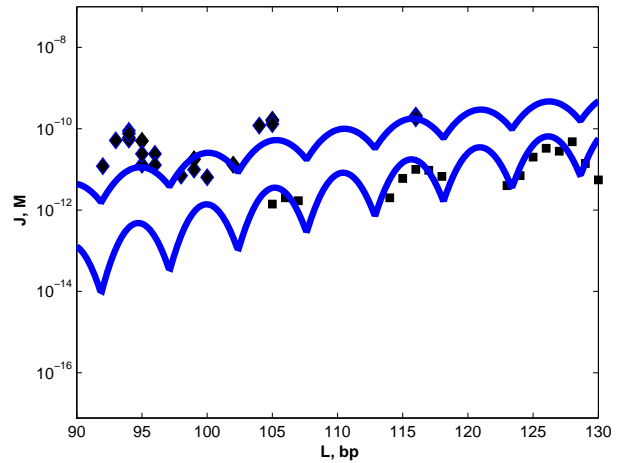}
\end{center}
\caption{Comparison of \ the theoretical result, Eq. (\protect\ref{J0}) with
experimental data from Refs. \protect\cite{Widom1} (diamonds) and \ 
\protect\cite{volo} (squares). The two curves correspond to weak ($\protect%
\kappa=1,\protect\kappa^{\prime}=2$) and strong ($\protect\kappa=\protect%
\kappa^{\prime}=20$) orientational couplings, respectively. }
\label{exper}
\end{figure}

To complete our calculation we now need to include effects of the torsional
constrain. For short loops, the corresponding effective Hamiltonian is the
sum of the \ elastic energy of the twisted chain and the torsional part of $%
H_{end}$:%
\begin{equation}
H_{tors}\left( \Delta \psi \right) \approx \frac{l_{t}}{2L}\left[ \Delta
\psi +2\pi \left( N-\frac{L}{h}\right) \right] ^{2}+\frac{\kappa ^{\prime
}\left( \Delta \psi \right) ^{2}}{2}
\end{equation}%
Here $N$ is an integer\textit{\ linking number,} and $h$ is helix repeat of
dsDNA. This leads to an additional torsion-related factor in the final
result:%
\begin{equation}
J=\frac{J_{\kappa }\left( L/l_{p}\right) }{\sqrt{\kappa ^{\prime }L/l_{t}+1}}%
\sum\limits_{N=-\infty }^{+\infty }\exp \left[ \frac{2\pi ^{2}\left(
N-L/h\right) ^{2}}{L/l_{t}+\kappa ^{\prime -1}}\right]  \label{J}
\end{equation}

The calculated $J$-factor is shown in Figure \ref{exper}, as a function of
the loop size. Remarkably, the two conflicting sets of experiments are both
consistent with the model. In particular Cloutier-Widom and Vologoidskii lab
data correspond to the regimes of weak ($\kappa =1$, $\kappa ^{\prime }=2$)
and strong ($\kappa =\kappa ^{\prime }=20$) terminal coupling, respectively.
As expected, the \ two coupling parameters, $\kappa $ and $\kappa ^{\prime }$
are strongly correlated. This variation in value of $\kappa $ is reasonable
since the effective rigidity of a DNA "nick" is likely to have an
exponential dependence on the local stacking energy. This strong dependence
of the cyclization probability on the local sequence should not be confused
with another sequence dependent effect associated with inhomogeneous
intrinsic curvature and bending modulus of the chain \cite{PTloop}. In order
to separate the two effects, the sequence of the mutually complementary
terminal groups and that of the rest of the chain should be varied
independently in the future experiments.

One can make several important conclusions based on our results. First, the
traditional "circular loop" modelling of DNA cyclization is only justified
for very strong coupling, $\kappa \gtrsim 10$. Second, in the regime of
moderate coupling ($\kappa \simeq $ $\kappa ^{\prime }\sim 1$) the effects
of the two constraints are rather different. The overall shape of $J$-factor
curve follows that of an orientationally unconstrained loop, Eq. (\ref{J1}),
but the torsion-related oscillations, although greatly reduced, remain
rather prominent. This is indeed consistent with the results of Ref. \cite%
{Widom2}.

\end{document}